\documentclass[preprint]{elsarticle}[10pt]
\usepackage{epsfig}
\usepackage{supertabular}
\usepackage{natbib}
\begin{document}

\title{A rapid algorithm to calculate joint probability matrices for joint entropies of arbitrary order}

\author{Reginald D. Smith}
\address{Citizen Scientists League \\ PO Box 10051 \\Rochester, NY 14610}
\ead{rsmith@citizenscientistsleague.com} 
\cortext[cor1]{Corresponding author}
\date{November 25, 2014}

\begin{abstract}
There is no closed form analytical equation or quick method to calculate probabilities based only on the entropy of a signal or process. Except in the cases where there are constraints on the state probabilities, one must typically derive the underlying probabilities through search algorithms. These become more computationally expensive as entropies of higher orders are investigated. In this paper, a method to calculate a joint probability matrix based on the entropy for any order is elaborated. With this method, only first order entropies need to be successfully calculated while the others are derived via multiplicative cascades.
\end{abstract}


\maketitle

Entropy, a common variable describing signals and processes, is based on the underlying single or joint probabilities. However, calculating these probabilities from only the value of entropy is extremely difficult except in trivial cases such as uniform distributions which give a maximum entropy based on the number of symbols ($\log M$) or situations where certain maximum entropy principles are used and constraints on the probabilities are known beforehand to eliminate most possible solutions \cite{jaynes} .

In this paper, a quick method for calculating the joint probability matrices of any order is elaborated. The conditional entropy, the difference between the joint entropy being calculated and the previous order's single or joint entropy, is used to create a probability vector which can be used the derive the next order's joint probabilities. While probabilities representing first order entropies need to be repeatedly calculated at each step, the overall joint probability is then derived from the previous order's joint probability matrix via a multiplicative cascade.

\section{Derivation}

For example, take the second order joint probability $p(x_i,x_j)$.

\begin{equation}
p(x_i,x_j)=p(x_j|x_i)p(x_i)
\end{equation}

Where $p(x_j|x_i)$ is the probability of symbol $j$ given the occurrence of symbol $i$. As a simplifying assumption, assume that the probability of $j$ occurring after any symbol $i$, is identical for all $i$ so that $p(x_j|x)=p(x_j^*)$ for all $i$ and where $p(x_j^*)$ is the conditional probability for $j$ given any preceding symbol. By definition we also have 
\begin{equation}
p(x_j)=\sum_{i=1}^N p(x_i)p(x_j^*)
\end{equation}

where $N$ is the total number of symbols with probability. If all symbols have a similar conditional probabilities, the joint entropy 

\begin{equation}
H(x_i,x_j)=-\sum_{i} \sum_{j} p(x_i,x_j) \log p(x_i,x_j).
\end{equation}

can be re-defined as

\begin{equation}
H(x_i,x_j)=-\sum_{i} \sum_{j} p(x_i)p(x_j^*) \log p(x_i)p(x_j^*)
\end{equation}

Rearranging the equations can give another expression below

\begin{equation}
H(x_i,x_j)=-\sum_{i}p(x_i) \sum_{j}p(x_j^*) \log p(x_j^*) -\sum_{j}p(x_j^*) \sum_{i}p(x_i) \log p(x_i).
\end{equation}

Finally, since $\sum_{i}p(x_i)=\sum_{j}p(x_j^*)=1$ we can simplify the joint entropy to

\begin{equation}
H(x_i,x_j)=H(x) + H(x^*)
\label{simpjoint}
\end{equation}

In the above $H(x)$ is the first order entropy while $H(x^*)$ is a first order entropy with the same value as the conditional entropy and is defined as 

\begin{equation}
H(x^*) = -\sum_{j=1}^N p(x_j^*) \log p(x_j^*)
\end{equation}

This definition of joint entropy allows us to rewrite the joint probabilities as

\begin{equation}
 p(x,x^*)=p(x)p(x^*)
\end{equation}

Finally, equation \ref{simpjoint} can be again re-written as a joint entropy between $x$ and $x^*$

\begin{equation}
H(x_i,x_j)=H(x,x^*)=-\sum \sum p(x,x^*) \log p(x,x^*)
\end{equation}

\section{Calculating the joint probability matrix}

Given the single entropy or joint entropy of any order, one can calculate a joint probability matrix of any entropy through the following methodology. 

First, calculate the conditional entropy which is the joint entropy of the order to be calculated minus the joint entropy of the previous order.

Second, using the base number of symbols from the first order probabilities, calculate a probability vector with the same number of probabilities whose sum is 1 but whose first order entropy matches the conditional entropy from the first step.

\begin{figure}
\centering 
\includegraphics[width=5in, height=3in]{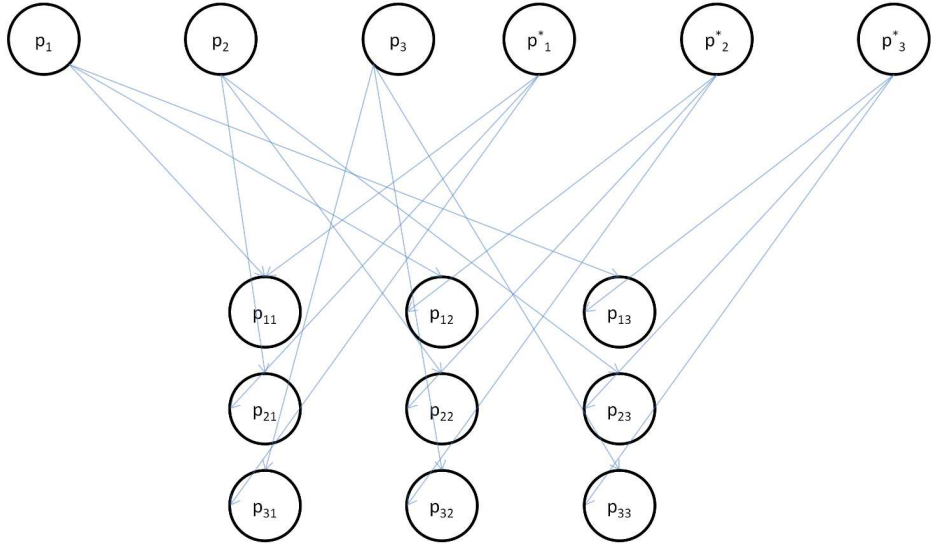}
\caption{Illustration of the multiplicative cascade process using three symbols. The symbols for the first order probabilities are $p_1$, $p_2$, and $p_3$ and the probabilities of the conditional entropy probability vector are $p_1^*$, $p_2^*$, and $p_3^*$.}
\label{cascadepic}
\end{figure}

Finally, use these probabilities as a multiplicative cascade and multiply each joint entropy times every one of these derived probabilities. The result will be a new joint probability matrix whose entropy matches the targeted joint entropy. Figure \ref{cascadepic} gives a graphic illustration of the cascade and multiplication process.

A quick example would be if you have a process which uses 10 distinct symbols. So the zeroth order entropy is $\log_2 10$=3.32 bits. Assume the first order is 2.5 bits and we want to calculate the second order joint probability matrix (with 100 joint probabilities) that has an entropy of 3.2 bits. The conditional entropy is 0.7 bits (3.2 bits for the second order minus 2.5 bits for the first order). 

First, we must derive the probability vector of the 10 symbols of the first order entropy. This part is best done using search algorithms, but is relatively fast and results in an unordered list of 10 probabilities summing to 1 with an entropy of 2.5 bits.

Second, we must derive the probability vector for $H(x^*)$ which are 10 probabilities who sum to 1 but whose entropy is 0.7 bits.  The resultant probabilities are our values for $p(x^*)$. This should be solved using a similar, rapid search algorithm as in the first step.

Finally, we must multiply each of these derived probabilities by each of the probabilities of the first order entropy. The result is a 100 entry joint probability matrix whose entropy is 3.2 bits. If we want to calculate a third order probability matrix (with 1,000 probabilities) with an entropy of 3.8 bits we follow the same methodology. Create a 10 symbol probability vector based on probabilities which sum to 1 but have a first-order entropy of 0.6 bits. Then multiply these derived probabilities by the second order joint probability matrix. The result will be a 1,000 element third order joint probability matrix of 3.8 bits joint entropy. This method can be extended for any arbitrary order of joint entropy with the same results.

\end{document}